\documentclass[apjl,twocolumn]{emulateapj}
\usepackage{epsfig,mathptmx,graphicx}

\def\gtsima{$\; \buildrel > \over \sim \;$}
\def\ltsima{$\; \buildrel < \over \sim \;$}
\def\prosima{$\; \buildrel \propto \over \sim \;$}
\def\gsim{\lower.5ex\hbox{\gtsima}}
\def\lsim{\lower.5ex\hbox{\ltsima}}
\def\simgt{\lower.5ex\hbox{\gtsima}}
\def\simlt{\lower.5ex\hbox{\ltsima}}
\def\simpr{\lower.5ex\hbox{\prosima}}

\def\h1{$h^{-1}$}
\def\eeq{\end{equation}}
\def\beq{\begin{equation}}

\shorttitle{Milky-Way like Molecular Gas Excitation of massive disk galaxies at $z\sim1.5$}
\shortauthors{Dannerbauer et al.}

\begin{document}


\title{Low,  Milky-Way like, Molecular Gas Excitation of Massive Disk Galaxies
at $\lowercase{z}\sim1.5$}


\author{H.~Dannerbauer\altaffilmark{1},
        E.~Daddi\altaffilmark{2},
	D.A.~Riechers\altaffilmark{3,8},
	F.~Walter\altaffilmark{1},
	C.L.~Carilli\altaffilmark{4},
	M.~Dickinson\altaffilmark{5},
	D.~Elbaz\altaffilmark{2},
	G.E.~Morrison\altaffilmark{6,7}
}

\altaffiltext{1}{Max-Planck-Institut f\"ur Astronomie, K\"onigstuhl 17, D-69117 Heidelberg, Germany; dannerb@mpia-hd.mpg.de}
\altaffiltext{2}{CEA, Laboratoire AIM,
    Irfu/SAp, Orme des Merisiers, F-91191 Gif-sur-Yvette, France}
\altaffiltext{3}{Caltech, MC 249-17, 1200 East California Boulevard, Pasadena, CA 91125}
\altaffiltext{4}{NRAO, P.O. Box O, Socorro, NM 87801-0387}
\altaffiltext{5}{NOAO,  950 N. Cherry Ave., Tucson, AZ, 85719}
\altaffiltext{6}{IfA, University of Hawaii, Honolulu, HI 96822}
\altaffiltext{7}{CFHT, Kamuela, HI 96743}
\altaffiltext{8}{Hubble Fellow}








\begin{abstract}
We present evidence for Milky-Way-like, low-excitation molecular gas
reservoirs in near-IR selected massive galaxies at $z\sim1.5$, based
on IRAM Plateau de Bure Interferometer CO[3-2] and NRAO Very Large
Array CO[1-0] line observations for two galaxies that had been
previously detected in CO[2-1] emission.  The CO[3-2] flux of
$BzK-21000$ at $z=1.522$ is comparable within the errors to its
CO[2-1] flux, implying that the CO[3-2] transition is significantly
sub-thermally excited.  The combined CO[1-0] observations of the two
sources result in a detection at the $3\sigma$ level that is
consistent with a higher CO[1-0] luminosity than that of CO[2-1].
Contrary to what is observed in submillimeter galaxies and QSOs, in
which the CO transitions are thermally excited up to $J\geq3$, these
galaxies have low-excitation molecular gas, similar to that in the
Milky Way and local spirals.  This is the first time that such
conditions have been observed at high redshift.  A Large Velocity
Gradient analysis suggests that
molecular clouds with density 
and kinetic temperature 
comparable to local spirals can reproduce our observations.  
The similarity in the CO excitation properties suggests that a high,
Milky-Way-like, CO to H$_{2}$ conversion factor could be appropriate
for these systems.  If such low-excitation properties are
representative of ordinary galaxies at high redshift, centimeter
telescopes such as the Expanded Very Large Array and the longest
wavelength Atacama Large Millimeter Array bands will be the best tools
for studying the molecular gas content in these systems through the
observations of CO emission lines.
\end{abstract}





\section{Introduction}
Daddi et al. (2008; D08 henceforth) recently reported the discovery of
large molecular gas reservoirs inside near-IR selected galaxies with
star formation rates $SFR\sim 100-150~M_{\sun}~yr^{-1}$ and stellar
masses $M^*\sim10^{11}M_\odot$ at redshift $z\sim1.5$.  The SFR in
these systems is an order of magnitude lower than in submillimeter
galaxies (SMGs) and QSOs that were also previously detected in
molecular gas emission at high redshift \citep[e.g.,][]{gre05,rie06}.
The CO[2-1] luminosities derived from IRAM Plateau de Bure
Interferometer (PdBI) observations of two sources --- $BzK-4171$ and
$BzK-21000$ at $z=1.465$ and $z=1.522$ respectively --- imply
star-formation efficiencies (SFEs) similar to those of local spirals,
an order of magnitude lower than those in submillimeter selected
sources.  The low SFEs suggest that these objects behave as scaled-up
versions of local spirals, with long lasting, low-efficiency
star-formation.  Given the disk-like UV rest-frame morphology of these
objects in deep HST+ACS imaging, and the good agreement between the UV
and radio/mid-IR star formation rates, D08 suggested that star
formation in these galaxies is taking place in disk-wide extended gas
reservoirs. This is contrary to major merger-triggered nuclear
starbursts as seen in local ULIRGs and distant SMGs, despite the fact
that the SFRs and star formation surface densities in these objects
are similar to those of local starbursts and ULIRGs.  Given that both
galaxies observed in CO by D08 were detected with large CO
luminosities, and that the space density of similar sources is
10--30$\times$ larger than that of SMGs (Daddi et al. 2007), this
suggests that such a quiescent gas-consumption activity could be a
prevalent mode of star-formation and galaxy growth in the distant
universe.

In order to gain insights on the physical properties of the molecular
gas and thus on the nature of the star formation in these galaxies it
is important to investigate the CO excitation properties. In galaxies
like the Milky Way \citep{fix99} and local star-forming spiral
galaxies \citep[e.g.,][]{bra92,you95,mau99,cro07,pap98} the molecular
gas has relatively low excitation and is rather diffusely distributed
(low density and temperature), resulting in a high CO to H$_{2}$
conversion factor.  In contrast, dusty starburst systems like ULIRGs
in the local Universe and SMGs and QSOs in the distant universe have
very dense and warm gas \citep[e.g.,][]{tac06,tac08,wei05a,wei05b,rie06,dow98}, with
a CO spectral energy distribution showing thermalized emission up to
at least the rotational transition $J=3$ \citep{wei07b}, and a CO to
H$_{2}$ conversion factor about 5 times lower than in typical spirals
\citep{sol05}.  Despite many CO observations at high redshift,
no evidence for Milky-Way-like CO excitation has been found so far.
This result may be biased by the pre-selection of the most IR-luminous
galaxies for study, and by the prevalence of high frequency, high-J
transition observations.

Here we present multi-J CO observations based on PdBI and Very Large
Array (VLA) observations for two $z\sim1.5$ BzK-selected galaxies
previously detected in CO[2-1] with the aim to estimate the excitation
properties of CO in these sources.

\section{Observations}
\label{sec:observations}
\subsection{Plateau de Bure Observations}
\label{sec:pdbiobs}

We observed $BzK-21000$ at $z=1.522$ in the CO[3-2] transition
(rest-frame 345.796~GHz) with the PdBI in D-configuration in May and
June 2008. The on-source observing time was 9.7~hours with the full
array (six antennae) and a beam size of $4.0\arcsec\times3.1\arcsec$.
We tuned the receivers to 136.970~GHz (2mm window), slightly offset to
the frequency of the CO[3-2] line at $z=1.522$ at 137.112~GHz.  The
total bandwidth of our dual polarization mode observations was 1~GHz
(corresponding to $\sim$2200~km~s$^{-1}$).  The correlator has 8
independent units each covering 320~MHz (128~channels each with a
width of 2.5~MHz). The phase center was 7.5\arcsec\/ away from the
position of $BzK-21000$ in order to target the SMG GN20 during the
same observations (Daddi et al. 2009). All the values quoted below
were corrected for a primary beam attenuation (PBA) of 11\%.  We
reduced and calibrated the data with the GILDAS software packages CLIC
and MAP. The final data, imaged using natural weighting, have a noise
of 0.11~mJy integrated over the full 1~GHz bandwidth.  In addition to
the CO[2-1] dataset already presented by D08, we also use CO[2-1]
observations of both BzK galaxies obtained with IRAM PdBI in
B-configuration (synthesized beam $\sim1.3''$) that have been already
presented in Daddi et al. (2009). The new data solidly confirm the
CO[2-1] detections of D08. Our flux calibration in both the 2mm and
3mm observations with IRAM PdBI is primarily based on measurements of
the standard calibrator $MWC349$ for which accurate models of the
spectrum at these frequencies are available. The knowledge of the
typical ranges of PdBI antennas efficiencies at the various
frequencies can also be used as a cross check of the flux
calibration. Overall, we estimate that the typical accuracy on the
absolute flux calibration for our observations are within a range of
$\sim$10\% in the 3mm band and $\sim$22\% in the 2mm band. To the
measurement errors, we add in quadrature uncertainties corresponding
to half those ranges for the estimate of the total error budget on
fluxes and luminosities in each observing transition.
 

\begin{figure}
\begin{center}
\includegraphics[angle=270,scale=.575]{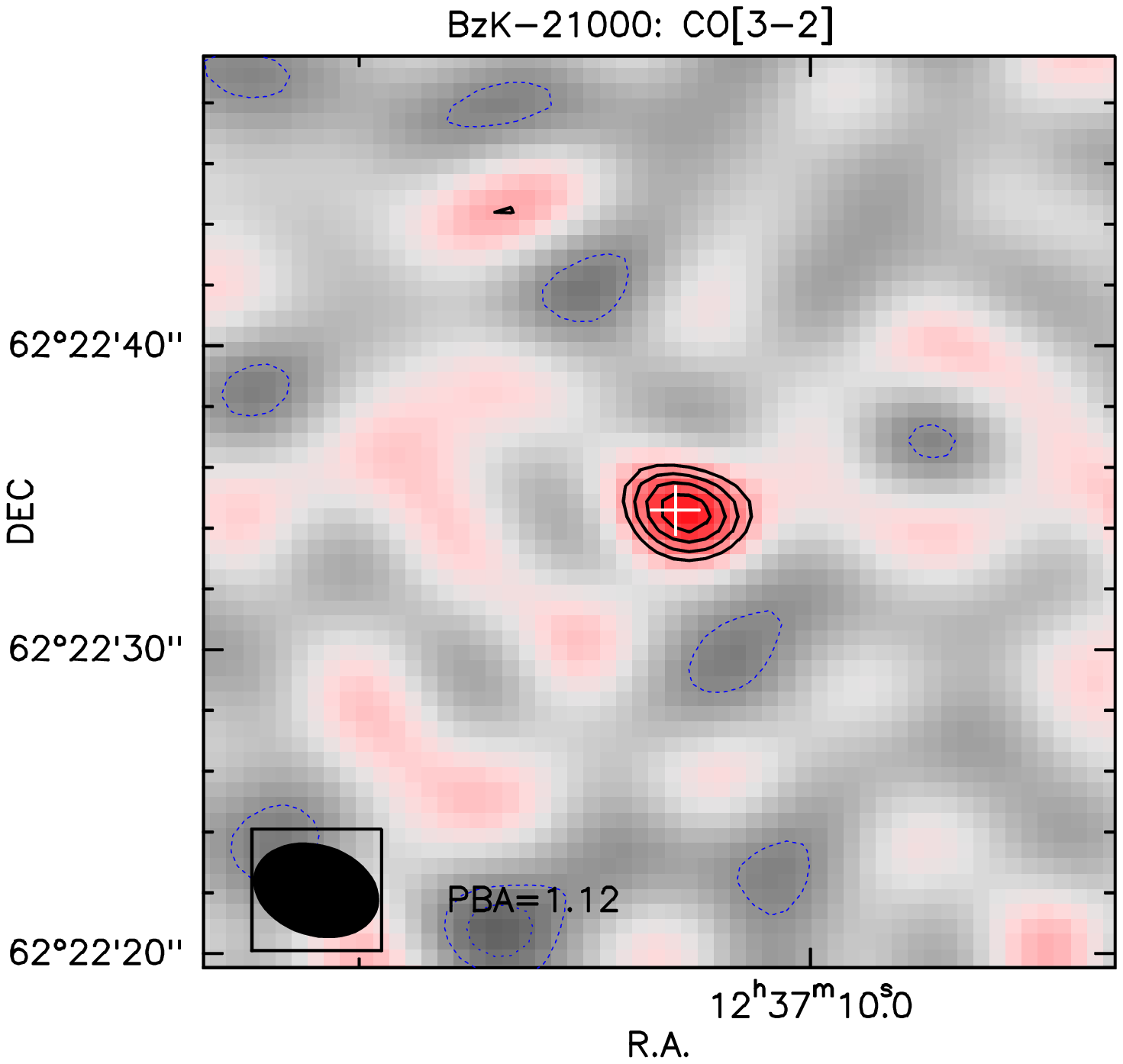}
\includegraphics[angle=270,scale=.575]{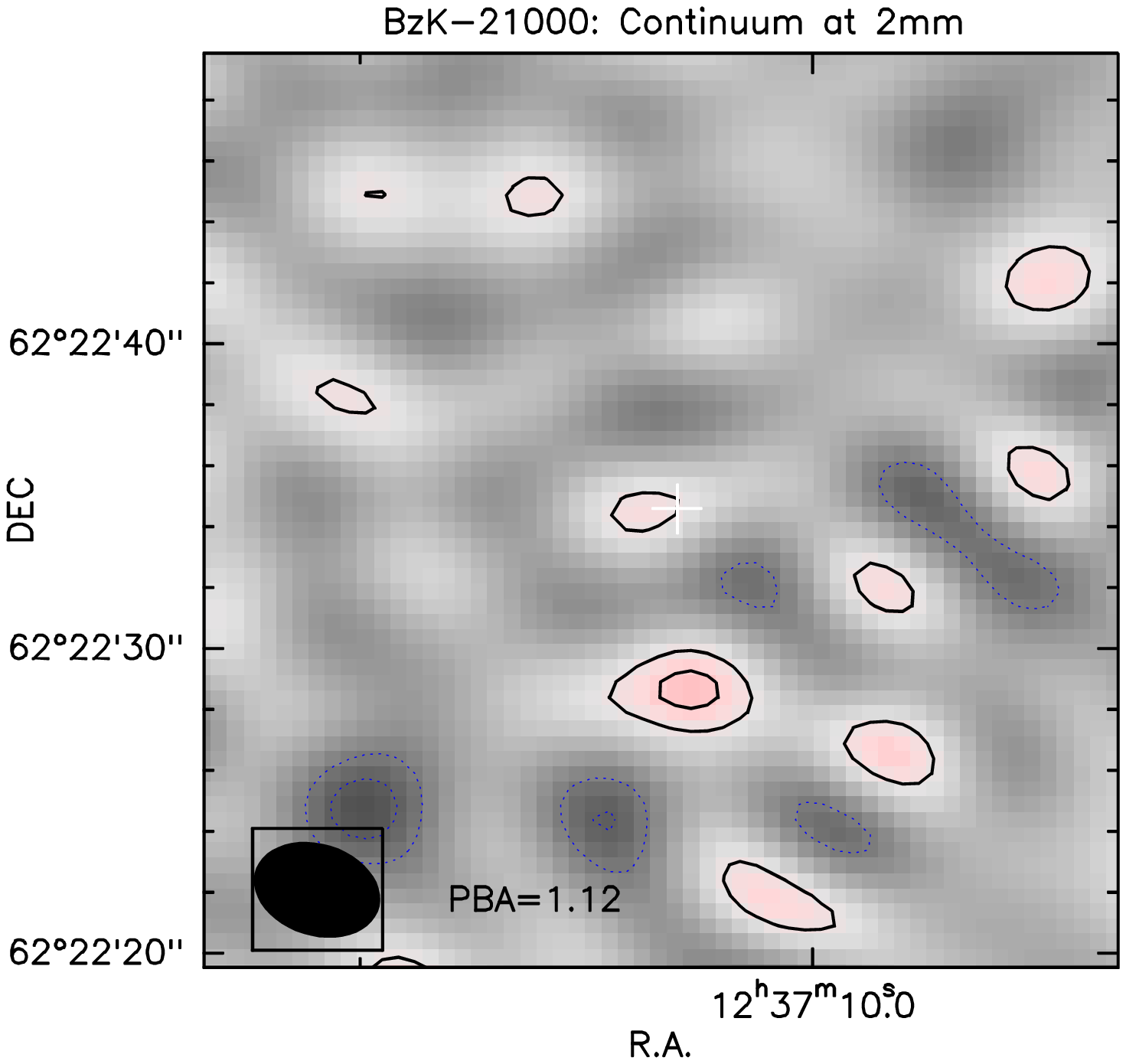}
\caption{Top panel: CO[3-2] map 
centered on $BzK-21000$, averaged 
over
525~km~s$^{-1}$, corresponding to the observed range of CO emission
(see also Fig.~\ref{fig:spectra}). Contour levels start at $\pm2\sigma$
and are 
in steps of $1\sigma$ ($\sim0.215$~mJy~beam$^{-1}$) with
positive (negative) contours shown as solid (dashed) lines. 
The white cross shows the VLA 1.4~GHz position. Bottom panel: 
map averaged over the remaining channels.
}\label{fig:co32}
\end{center}
\end{figure}

\subsection{Very Large Array Observations}
\label{sub:vla}

We observed both sources --- $BzK-21000$ and $BzK-4171$ --- with the
VLA in D-configuration in August and September 2008 (synthesized beam
of $\sim1.5\arcsec$). In order to measure the CO[1-0] transition
(rest-frame 115.271~GHz) we used two contiguous 50~MHz IF bands tuned
at 45.6851~GHz and 45.7351~GHz for $BzK-21000$ and at 46.7351~GHz and
46.7851~GHz for $BzK-4171$. The total bandwidth of 100~MHz corresponds
to about 650~km~s$^{-1}$ in velocity space, which covers well most of
the emission lines. The total integration time was 21~h and 28~h
on-source for $BzK-21000$ and $BzK-4171$, respectively.  We reduced
and calibrated the data with standard techniques using AIPS. Given
that summer time observations with the VLA can be subject to bad phase
stability, we employed fast switching phase calibration using the
calibrator J1302$+$578. We also included cycles where we switched to a
position near to the phase calibrator, and used these data to check
the phase coherence time.  From this we measured the flux correction
factor due to de-coherence of 24\% for $BzK-21000$ and 15\% for
$BzK-4171$. We correct our measurements by these factors to account
for these small losses of signal.  The final maps (100~MHz bandwidth)
have rms noise levels of 0.11~mJy and 0.13~mJy for $BzK-21000$ and
$BzK-4171$, respectively, once corrected for de-coherence and PBA at
the 10\% level. For the VLA observations, by monitoring the
bootstrapped flux of the phase calibrator over the observing days, we
estimate a gain calibration uncertainty in a range of 14\%, small
compared to the typical measurement uncertainties for our objects (see
next section).

\section{Results}
\subsection{CO[3-2] in $BzK-21000$}

\begin{figure}
\begin{center}
\includegraphics[scale=.44]{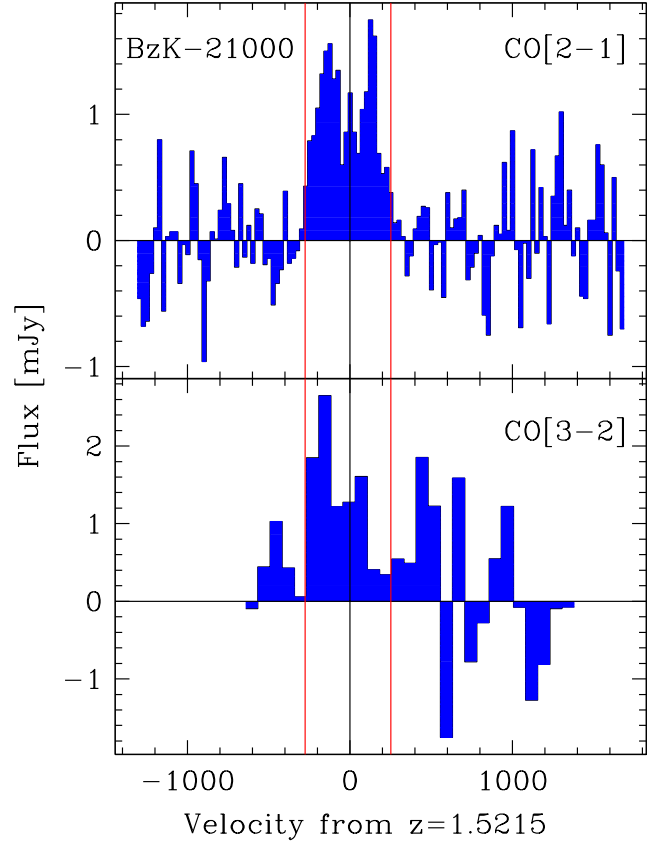}
\caption{The CO spectra of $BzK-21000$. CO[2-1] is in the {\it
 top panel} (sampling of 25~km~s$^{-1}$) and CO[3-2] is in the {\it bottom panel} (sampling of  75~km~s$^{-1}$). 
 The red horizontal line indicates the
 velocity range used for measuring the 
CO fluxes.
The spectra are continuum 
subtracted as discussed in the text.}\label{fig:spectra}
\end{center}
\end{figure}

The CO[3-2] transition in $BzK$-21000 is securely detected in the IRAM
PdBI data at the $\sim6\sigma$ level (Fig.\ref{fig:co32}).  In order
to properly compare the fluxes from the different CO transitions, and
to derive a CO spectral line energy distribution (SLED), we need to
perform flux measurements over the same velocity range in the
different transitions.  We use a velocity range of 525~km~s$^{-1}$
centered on the CO[2-1] line (the transition detected with the highest
S/N ratio, Fig.~\ref{fig:spectra}).  We also corrected line fluxes for
possible underlying continuum emission. Direct measurements in the
PdBI maps return no significant evidence for continuum emission
(Fig.~\ref{fig:co32} and~\ref{fig:spectra}) with formal measurements
of $S({\rm 3mm})=40\pm60\mu$Jy and $S({\rm
2mm})=320\pm150\mu$Jy. Given the estimate of $L_{\rm
IR}\sim10^{12}L_\odot$ in these galaxies (D08) we would expect
continuum emission at the level of $S({\rm 3mm})=30\mu$Jy and $S({\rm
2mm})=100\mu$Jy based on the SED models of Chary \& Elbaz (2001) that
have been shown to accurately reproduce the properties of BzK galaxies
(e.g., Daddi et al. 2005; 2007). These predictions are consistent with
our measurements.  Given the non-detections of the continuum in our
observations, we used the model estimates as our best guesses and
subtracted those values to obtain continuum-free line
measurements. These continuum corrections are small, 7\% and 2.5\% for
the 2mm and 3mm band, respectively. We did not correct the VLA
measurements for continuum emission, which is expected to be entirely
negligible at 45~GHz ($\ll 10~\mu$Jy).

Applying all the corrections and adding all sources of uncertainty in
quadrature, our CO measurements for $BzK-21000$ are $I_{\rm
CO[3-2]}=0.70\pm0.15$~Jy~km~s$^{-1}$ and $I_{\rm
CO[2-1]}=0.62\pm0.07$~Jy~km~s$^{-1}$.  This new CO[2-1] measurement is 
slightly lower than that reported in D08 but it has a higher S/N ratio.
The  CO[2-1] spectrum now clearly shows a double peak profile, similar to $BzK-4171$ in D08,
suggestive of rotation.

In the following we express the line ratios in terms of brightness
temperature ratios (or equivalently, luminosity ratios) of the
different transitions, as can be derived from equation 3 of Solomon \&
van den Bout 2005): $r_{J\,
J-1}=I_{CO[J-(J-1)]}/I_{CO[(J-1)-(J-2)]}*((J-1)^2/J^2)$.  In the case
of thermalized transitions the temperature ratio is 1 by definition,
the CO luminosity is the same in the different transitions and the CO
fluxes in Jy~km~s$^{-1}$ scale proportionally to $J^2$.

For $BzK-21000$ we derive a solid measurement of $r_{32}=0.50\pm0.12$.
The $r_{32}$ measurement clearly shows that the CO[3-2] transition is
sub-thermalized with respect to CO[2-1], a result significant at the
$>4\sigma$ level. From Fig.~\ref{fig:spectra} it seems that
CO[3-2] might be particularly weak in the reddest half of
the spectrum, a possible sign of differential excitation. 
Measuring $r_{32}$
for the red and blue half separately we do find a suggestion for
different excitation ($r_{32}=0.28$ and 0.64, respectively), but the
difference is significant only at the 1.7$\sigma$ level.

\begin{figure}
\begin{center}
\includegraphics[scale=.54,angle=-90]{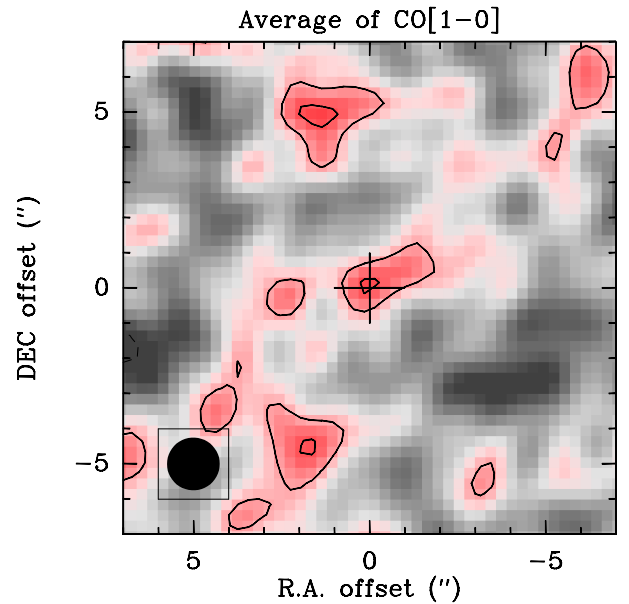}
\caption{Average map of CO[1-0] VLA observations of $BzK-21000$ and
$BzK-4171$, covering 650~km~s$^{-1}$ well centered on both lines. 
Contours start at $\pm2\sigma$. The map has a noise of 90$\mu$Jy.
The cross corresponds to the VLA 1.4GHz continuum positions.
}
\label{fig:vla}
\end{center}
\end{figure}

\begin{figure*}
\begin{center}
\includegraphics[angle=0,scale=.46]{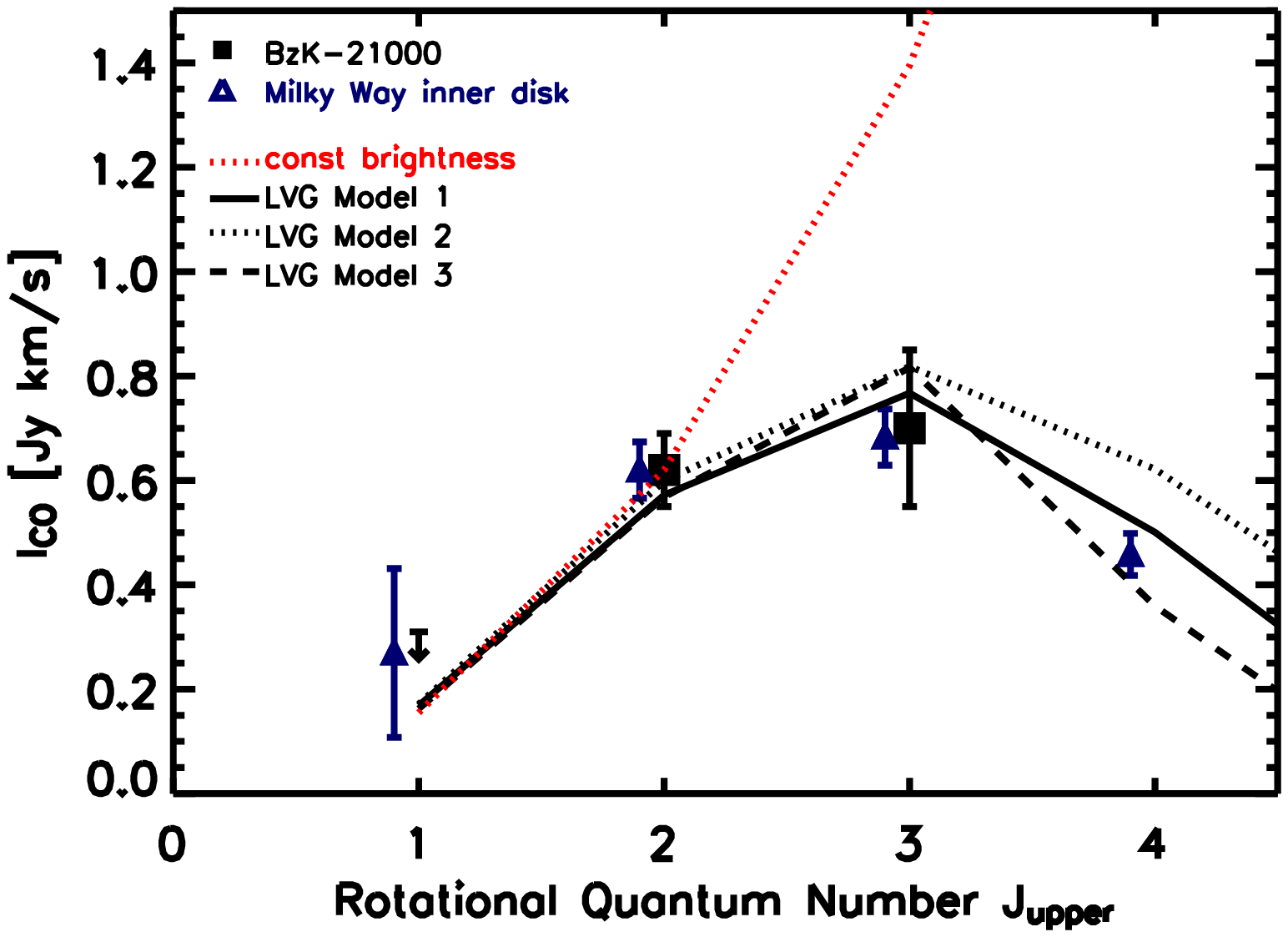}
\includegraphics[angle=0,scale=.46]{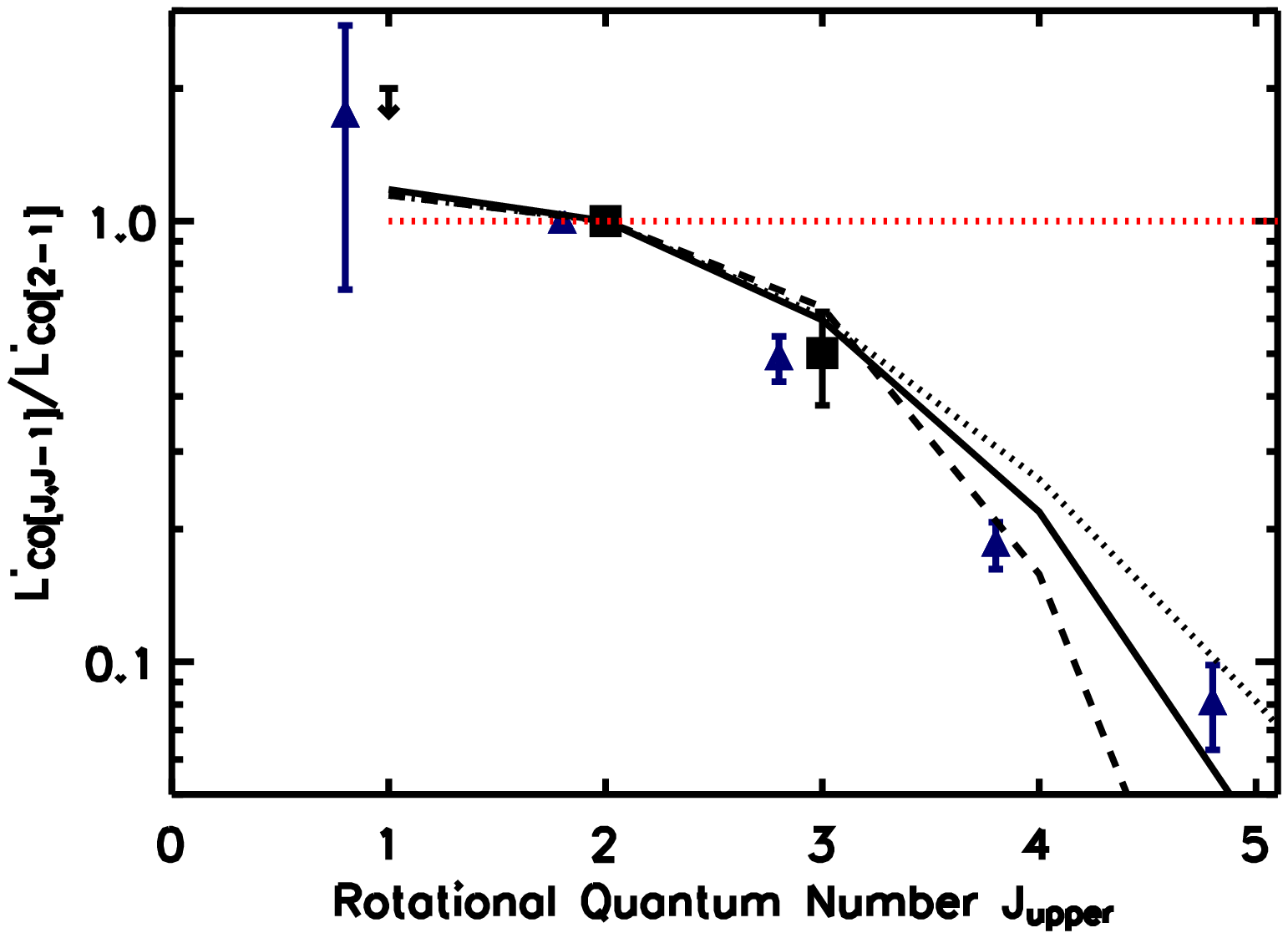}
\caption{Left panel: The CO 
SLED of
$BzK-21000$ at $z=1.522$ obtained from PdBI and VLA observations. The
black solid, dotted and dashed lines show our LVG models (model 1, 2, 3, respectively).
The Milky Way SLED from 
\citet[triangles,][]{fix99} is shown for comparison.  The constant brightness temperature
relation (red dashed line) is normalized to our CO[2-1] measurement. Right panel: The same of left panel but shown in luminosity units, normalized to the CO[2-1] transition. $J_{upper}>2$ transitions can severely underestimate the CO
luminosity and thus the molecular gas mass. 
}\label{fig:cosed}
\end{center}
\end{figure*}

\subsection{CO[1-0] constraints}

Inspection of the VLA observations of the two BzK galaxies shows
$\simgt2\sigma$ peak signal close to the positions of both galaxies.
Individual fluxes cannot be reliably derived from the data.  In order
to increase the S/N in the measurements we corrected each VLA map for
de-coherence and PBA, aligned the maps at the VLA 1.4~GHz continuum
coordinates of the two galaxies and averaged the data from the two
objects.  Fig.~\ref{fig:vla} shows that some positive CO[1-0] signal
is detected at the expected position, significant at about 3.1$\sigma$
and corresponding to a peak flux density of $0.28\pm0.09$mJy.  Given
the insufficient S/N of the data, it is impossible to know if the
CO[1-0] emission is slightly extended.  In order to compare this
CO[1-0] measurement with the CO[2-1] results we have used the
B-configuration CO[2-1] observation with IRAM PdBI, which have
comparable angular resolution to the VLA CO[1-0] data, and extracted
the signal over velocity ranges matching the VLA observations on each
galaxy (covering very well both lines). This results on an average
peak CO[2-1] signal in the combined CO[2-1] map of $0.67\pm0.09$~mJy.
We derive $r_{21} = 0.60^{+0.40}_{-0.17}$, averaged over the two BzK
galaxies.  Taken at face value, this implies that we are seeing excess
peak emission in CO[1-0] relative to what is expected from CO[2-1]
although the evidence for an excess is very mild given the low S/N in
the CO[1-0] data.  Higher S/N observations would be needed to
establish if the CO[2-1] transition is sub-thermally excited with
respect to CO[1-0].

\section{Discussion}
\label{sec:discussion}

For $BzK-21000$, the CO[3-2] and [2-1] fluxes are comparable within
the errors.  The CO spectral line energy distribution of $BzK-21000$
is consistent with the Milky-Way disk measurements by \citet{fix99}
(Fig.~\ref{fig:cosed}), and similar to what is observed in local
spirals \citep[e.g.,][]{bra92,you95,mau99,cro07}.  The CO SLED of
$BzK-21000$ is in stark contrast to what has been generally seen so
far in high-redshift sources like QSOs and SMGs, for which gas has
been found to be thermalized up to at least the $J=3$ transition
\citep{wei05b,wei07a,wei07b,rie06}.  A possible exception is the
extremely red galaxy HR~10 at $z=1.44$ according to the results of
\citet{and00} and \citet{pap02} who find comparable fluxes of CO[5-4]
and CO[2-1].  However we note that the excitation in our sources is
still lower than that in HR10.

We now discuss the implication of our excitation measurements on the
physical properties of the gas, including the kinetic temperature
($T_{kin}$) and density ($n$) of the H$_{2}$ gas in the molecular
clouds, by using Large Velocity Gradient (LVG) modeling
\citep{gol74,sco74b}.  
We use the collision rates from \citet{flo01}
with a fixed CO abundance to velocity gradient ratio of
[CO]/(dv/dr)$=1\times10^{-5}~{\rm pc}~({\rm km}~{\rm s}^{-1})^{-1}$
{\citep[][]{wei05a}}, an H$_{2}$ ortho:para ratio of 3:1 and the cosmic
background temperature at $z=1.522$ of 6.89~K\ \footnote{Even for the
lowest CO[1-0] transition the cosmic background radiation at $z =
1.5$ will have only a minor effect on the depopulation of the state of
our two sources. The reason is that the intrinsic temperature in the
molecular clouds is expected to be higher, e.g. the intrinsic
temperature within the Milky-Way clouds is at least greater than 10~K
and most likely lies around $\sim20~$K.}.  
Due to the limited amount of information
available, we will not be able to resolve the ambiguity between all the
free parameters. E.g., choosing a different value of [CO]/(dv/dr) would also affect
the SLED, and would require re-adjusting $T_{kin}$ and/or $n$ to reproduce observed
ratios \citep[see, e.g.,][for a more detailed discussion of the 
parameter space degeneracies]{wei05a}.
We present here three representative models,
spanning the range of parameter degeneracies in $T_{kin}$ and $n$ (see
Fig.~\ref{fig:cosed}), that reproduce the observed CO SED of
$BzK-21000$.  Model~1 has a kinetic gas temperature $T_{kin}=25$~K and
molecular cloud density of ${n}=1300$~cm$^{-3}$, very similar to the
molecular gas excitation conditions of the Milky-Way.  The filling
factor of clouds, assuming a disk radius $r_{0}\sim5$~kpc from our ACS
observations (D08), is 2\% for this model.  Model~2 can still fit the
data assuming $T_{kin}=90$~K and $n=600$~cm$^{-3}$ with a similar
filling factor to model~1.  Model~3 results in a higher filling factor
of 8\% with a low $T_{kin}=10$~K but a higher $n=2500$~{\rm
cm}$^{-3}$.  Typical gas/dust temperatures in nearby
spirals/starbursts are 10--30 K \citep[e.g.,][]{bra92,mau99},
sometimes up to 40-50 K in the central regions of nuclear starbursts
\citep{dow98}. Normal galactic giant molecular clouds have typical
average densities of $10^{2}-10^{3}$~cm$^{-3}$
\citep{sco74a,eva99}. These values are broadly similar to what is
allowed from LVG modeling of our observations. 
Clearly, we can rule
out solutions where the gas is at high density and/or temperature,
similar to what is generally found inside local ULIRGs, distant SMGs
and QSOs.

Despite the degeneracies in the inferred physical parameters, the LVG
models presented above consistently predict $r_{21}\sim0.85$, higher
than our average estimate for the BzK galaxies but consistent within
the uncertainties. Robustly confirming an excess CO[1-0] emission over
what is predicted by the LVG models shown in Fig.~\ref{fig:cosed}
would imply the discovery of very cold molecular gas unseen even in
the CO[2-1] transition. For the moment, given our limited S/N at the
measured $r_{21}$, we assume that $r_{21}\sim0.85$ and
$r_{32}\sim0.50$ are the typical values for our sources.  This implies
that the CO luminosities derived for our galaxies in D08 from CO[2-1]
do not need substantial revisions, exception made for the small 16\%
increase required when converting CO[2-1] to CO[1-0], as suggested by
our LVG modeling.  However, our results strongly indicate that when
using transitions higher than CO[2-1] only a fraction of 
the CO[1-0] luminosity
is recovered (Fig.~\ref{fig:cosed}, right), implying that
such measurements are not reliable for accurately estimating the total
gas mass and star formation efficiencies in BzK-like galaxies. This
evidence could explain the discrepancy between the observations of
Hatsukade et al. (2009) and Tacconi et al. (2008) who failed to detect
CO[3-2] in 4 optically and UV selected galaxies at $1.4<z<2.5$, while
so far we have a high detection rate in CO[2-1] at redshift
$z\sim1.5$.  A large correction factor of 2.4 is required, in fact, at
CO[3-2] in order to recover the total molecular gas mass, based on our
results from $BzK-21000$.

Given that these two high redshift galaxies have star-formation
efficiencies and CO excitation properties similar to those in local
spirals and in the Milky Way, it is plausible that they may also have
a Milky-Way-like molecular conversion factor $\alpha_{CO}$ as well.
This would imply molecular gas reservoirs of $\sim10^{11}~M_{\sun}$
and gas mass fractions $f_{gas}\ga0.6$.  To summarize, our sources
provide clear examples of the long-sought after population of
low-excitation CO emitters at high-z.  Observations of a larger sample
of near-IR selected galaxies are required in order to confirm our
results in a statistical way. In particular, high S/N observations of
CO[1-0] and high-J CO are required to search for additional components
of colder and warmer gas, respectively.  Much higher S/N data will
also be required to search for excitation variations inside the disks,
as tentatively suggested by our data and as could be expected in the
case of very clumpy gas distribution (e.g., Bournaud et al. 2008).

Given that these are the first excitation measurements inside normal,
near-IR selected disk-like galaxies in the distant universe, it is
plausible to speculate that such low-excitation properties might be
typical in distant galaxies.  ALMA will be a powerful device to study
CO[2-1] in $z<3$ galaxies, but at much higher redshift, the higher
order transitions may not be excited in the average galaxy \citep[see
also][]{pap02}. In those cases, centimeter telescopes, such as the
EVLA or the Square Kilometer Array (SKA), will become the primary
tools for study of molecular gas in the earliest, normal galaxies.  In
distant galaxies with massive amounts of cold, diffuse, low-excitation
gas we will be able to detect carbon lines with ALMA
(e.g. [CII]$\lambda$158~$\mu$m redshifted to $>400\mu$m in $z>1.5$
galaxies).  This synergistic combination of observations will offer a
powerful tool for interpreting the gaseous content of ordinary
galaxies at high redshift.

\acknowledgments
Based on observations with the IRAM Plateau de Bure
Interferometer. IRAM is supported by INSU/CNRS (France), MPG (Germany)
and IGN (Spain).  The Very Large Array is a facility of the National
Radio Astronomy Observatory, operated by Associated Universities,
Inc., under a cooperative agreement with the National Science
Foundation.  We acknowledge the use of GILDAS software
(http://www.iram.fr/IRAMFR/GILDAS). We would like to thank Adam Leroy
for instructive conversations on the THINGS sample.  We thank Padeli
Papadopoulos, Axel Wei{\ss} and Frederic Bournaud for discussions and
Christian Henkel for providing the LVG code in its original
version. We thank an anonymous referee for useful comments that
improved our manuscript.  
ED and DE acknowledge the funding support of
French ANR under contracts ANR-07-BLAN-0228 and ANR-08-JCJC-0008.  DR
acknowledges support from NASA through Hubble Fellowship grant
HST-HF-01212.01-A awarded by the Space Telescope Science Institute,
which is operated by the Association of Universities for Research in
Astronomy, Inc., for NASA, under contract NAS 5-26555.

\clearpage





\begin{thebibliography}{}
\bibitem[Andreani et al.(2000)]{and00} Andreani, P., Cimatti, A., Loinard, L., R\"ottgering, H.\ 2000, \aap, 354, L1 




\bibitem[Bournaud et 
al.(2008)]{bour08} Bournaud, F., et al.\ 2008, \aap, 486, 741 

\bibitem[Braine 
\& Combes(1992)]{bra92} Braine, J., \& Combes, F.\ 1992, \aap, 264, 433 






\bibitem[Chary 
\& Elbaz(2001)]{cha01} Chary, R., \& Elbaz, D.\ 2001, \apj, 556, 562 

\bibitem[Crosthwaite 
\& Turner(2007)]{cro07} Crosthwaite, L.~P., \& Turner, J.~L.\ 2007, \aj, 134, 1827 

\bibitem[Daddi et al.(2005)]{dad05} Daddi, E., et al.\ 2005, 
\apjl, 631, L13 

\bibitem[Daddi et al.(2007)]{dad07} Daddi, E., et al.\ 2007, 
\apj, 670, 156 

\bibitem[Daddi et al.(2008)]{dad08} Daddi, E., Dannerbauer, 
H., Elbaz, D., Dickinson, M., Morrison, G., Stern, D., 
\& Ravindranath, S.\ 2008, \apjl, 673, L21 

\bibitem[Daddi et al.(2009)]{dad09} Daddi, E., et al.\ 2009, ApJ, 694, 1517



\bibitem[Downes 
\& Solomon(1998)]{dow98} Downes, D., \& Solomon, P.~M.\ 1998, \apj, 507, 615 

\bibitem[Evans(1999)]{eva99} Evans, N.~J., II 1999, \araa, 37, 311 

\bibitem[Fixsen et al.(1999)]{fix99} Fixsen, D.~J., Bennett, 
C.~L., \& Mather, J.~C.\ 1999, \apj, 526, 207 


\bibitem[Flower(2001)]{flo01} Flower, D.~R.\ 2001, Journal of 
Physics B Atomic Molecular Physics, 34, 2731 



\bibitem[Goldreich \& Kwan(1974)]{gol74} Goldreich, P., \& Kwan, J.\ 1974, \apj, 189, 441 

\bibitem[Greve et al.(2005)]{gre05} Greve, T.~R., et al.\ 
2005, \mnras, 359, 1165 




\bibitem[Hatsukade et al.(2009)]{hat09} Hatsukade, B., et 
al.\ 2009, PASJ, in press (astro-ph/0901.3388)


\bibitem[Mauersberger et al.(1999)]{mau99} Mauersberger, R., Henkel, C., Walsh, W., \& Schulz, A.\ 1999, \aap, 341, 256



\bibitem[Papadopoulos 
\& Seaquist(1998)]{pap98} Papadopoulos, P.~P., \& Seaquist, E.~R.\ 1998, \apj, 492, 521
 

\bibitem[Papadopoulos 
\& Ivison(2002)]{pap02} Papadopoulos, P.~P., \& Ivison, R.~J.\ 2002, \apjl, 564, L9 

\bibitem[Riechers et al.(2006)]{rie06} Riechers, D.~A., et 
al.\ 2006, \apj, 650, 604

\bibitem[Scoville et al.(1974)]{sco74a} Scoville, N.~Z., 
Solomon, P.~M., \& Jefferts, K.~B.\ 1974, \apjl, 187, L63 

\bibitem[Scoville \& Solomon(1974)]{sco74b} Scoville, N.~Z., \& Solomon, P.~M.\ 1974, \apjl, 187, L67 

\bibitem[Solomon \& Vanden Bout(2005)]{sol05} Solomon, P.~M., \& Vanden Bout, P.~A.\ 2005, \araa, 43, 677 



\bibitem[Tacconi et al.(2006)]{tac06} Tacconi, L.~J., et al.\ 
2006, \apj, 640, 228 

\bibitem[Tacconi et al.(2008)]{tac08} Tacconi, L.~J., et al.\ 
2008, \apj, 680, 246

\bibitem[Wei{\ss} et 
al.(2005a)]{wei05a} Wei{\ss}, A., Walter, F., \& Scoville, N.~Z.\ 2005a, \aap, 438, 533 

\bibitem[Wei{\ss} et 
al.(2005b)]{wei05b} Wei{\ss}, A., Downes, D., Walter, F., \& Henkel, C.\ 2005b, \aap, 440, L45 

\bibitem[Wei{\ss} et 
al.(2007a)]{wei07a} Wei{\ss}, A., Downes, D., Neri, R., Walter, F., Henkel, C., Wilner, D.~J., Wagg, J., \& Wiklind, T.\ 2007a, \aap, 467, 955 

\bibitem[Wei{\ss} et al.(2007b)]{wei07b} Wei{\ss}, A., Downes, D., 
Walter, F., 
\& Henkel, C.\ 2007b, From Z-Machines to ALMA: (Sub)Millimeter Spectroscopy of Galaxies, 375, 25 




\bibitem[Young et al.(1995)]{you95} Young, J.~S., et al.\ 
1995, \apjs, 98, 219 

\end{thebibliography}
\end{document}